\documentclass[amsmath, amssymb, twocolumn]{revtex4-1}
\draft

\newcommand{\nn}{\nonumber}

\newcommand{\intinf}[0]{\int\limits_{-\infty}^{+\infty}}
\newcommand{\eeqref}[1]{Eq.~(\ref{#1})}

\newcommand{\de}{\textrm{d}}

\usepackage{graphicx,dcolumn,bm,color,bbding,epstopdf}

\begin{document}

\title{Versatile Wideband Balanced Detector for Quantum Optical Homodyne Tomography}

\author{R.~Kumar}\affiliation{Institute for Quantum Information Science, University of Calgary, Calgary, Canada, T2N 1N4}

\author{E.~Barrios}\affiliation{Institute for Quantum Information Science, University of Calgary, Calgary, Canada, T2N 1N4}

\author{A.~MacRae}\affiliation{Institute for Quantum Information Science, University of Calgary, Calgary, Canada, T2N 1N4}

\author{E.~Cairns}\affiliation{Department of Chemistry, University of Calgary, Calgary, Canada, T2N 1N4}

\author{E.~H.~Huntington}\affiliation{School of Engineering and Information Technology,
University College, The University of New South Wales, Canberra ACT 2600, Australia}

\author{A.~I.~Lvovsky}\affiliation{Institute for Quantum Information Science, University of Calgary, Calgary, Canada, T2N 1N4}
\email{lvov@ucalgary.ca}

\date{\today}

\begin{abstract}
We present a comprehensive theory and an easy to follow method for the design and construction of a  wideband homodyne detector for time-domain quantum measurements. We show how one can evaluate the performance of a detector in a specific time-domain experiment based on electronic spectral characteristic of that detector. We then present and characterize a high-performance detector constructed using inexpensive, commercially available components such as low-noise high-speed operational amplifiers and high-bandwidth photodiodes. Our detector shows linear behavior up to a level of over 13 dB clearance between shot noise and electronic noise, in the range from DC to 100 MHz. The detector can be used for measuring quantum optical field quadratures both in the continuous-wave and pulsed regimes with pulse repetition rates up to about 250 MHz.
\end{abstract}


\maketitle

\section{Introduction}
The balanced homodyne detector (HD) is a useful tool in quantum optics and quantum information processing with continuous variables \cite{leo:97,lvov:09} since it can be used to measure field quadratures of an electromagnetic mode. These measurements provide information for complete reconstruction of quantum states in the optical domain (optical homodyne tomography).


With developing tools of continuous-wave quantum-optical state engineering \cite{bim:10} as well as state and process tomography \cite{lob:08}, the performance requirements for homodyne detectors continue to increase. The design of HDs for time-domain quantum tomography \cite{han:01, vogel:89, zav:02} is based on four main performance criteria: a) high bandwidth and a flat amplification profile within that bandwidth; b) high ratio of the measured quantum noise over the electronic noise; c) high common mode rejection ratio (CMRR); d) quantum efficiency of the photodiodes.

The \emph{high bandwidth} requirement comes from the fact that an HD must be able to measure field quadratures with sufficient time resolution. In the case of pulsed lasers, this corresponds to the inverse of the repetition rate of the pulses; in the case of a continuous signal, the required resolution is determined by the duration of the optical mode in which the signal states are produced \cite{sas:06, nie:09, yue:83}. This is technically challenging because most amplifiers have a limited gain-bandwidth product. Increasing the bandwidth implies reducing the gain, which, in turn, increases the effect of the electronic noise. Also the high frequency circuit layout poses a challenge to designers.

Within its bandwidth range, the HD must feature a \emph{flat amplification profile}. If this is not the case, the response of the HD to each individual pulse will exhibit ringing, which degrade the detector's time resolution and distort the measurement. This requirement also presents a major design challenge.

Any non-desirable ambient noises, dark current noises from the photodiodes and the intrinsic noise of the amplifiers fall under the umbrella of \emph{electronic noise}. The effect of this noise is to add a random quantity  $Q_e$ to the measurement of the field quadrature $Q_{\rm meas}$. This effect is equivalent to an additional optical loss channel with transmission \cite{appel:07}
\begin{equation}\label{etae}
 \eta_e=1-\langle\hat Q_e^2\rangle/\langle \hat Q_{\rm meas}^2\rangle.
\end{equation}
As we show below, the value of $\eta_e$ depends not only on the characteristics of the detector, but also on the conditions of the measurement in which the detector is used.

A HD must have a \emph{high subtraction capability} between the two photocurrents produced by the photodiodes. This can be expressed as a generalized common mode rejection ratio (CMRR) of the balanced detection \cite{vogel:89,chi:11}. The CMRR measures the ability of the device to reject the classical noise of the local oscillator \cite{bac:04,sas:06}.
This is particularly important in the pulsed case because a low CMRR (which implies a poor subtraction) will result in contamination of the signal with the repetition rate of the pulse and harmonics. Additionally, this lack of subtraction capability will make the HD more susceptible to saturation by the amplified signal from the photodiodes. High CMRR is difficult to achieve because the response functions of the photodiodes are not exactly the same. Therefore a pair of photodiodes with response functions as similar as possible must be chosen.

Experimentally, these performance benchmarks can be measured using an electronic spectrum analyzer. The spectrum of the homodyne output photocurrent gives information about the detector's bandwidth and amplification profile. Observing the output current in the absence of the local oscillator provides information about the magnitude and spectrum of the electronic noise. The lower bound on CMRR is determined by comparing the HD spectra when both photodiodes are illuminated and when only one is illuminated while the other is blocked.


In the present work, we quantitatively relate the measured electronic spectra to added noise in quadrature measurements. We show that the limited bandwidth and electronic noise can be translated into equivalent optical losses such as in \eeqref{etae}. We show how to estimate and reduce these losses for a specific time-domain experiment. In fact, in many cases (particularly, in the continuous-wave regime) electronic spectral measurements on the HD photocurrent in the presence and absence of the local oscillator are sufficient to precisely calculate the equivalent loss associated with the electronics.

The theoretical discussion in this paper is limited to the effects of the bandwidth and the electronic noise in two practically relevant regimes. The effect of the non-unitary quantum efficiency on quantum state reconstruction is well known \cite{leo:97}. A discussion of CMRR has been presented in detail in Ref.~\cite{chi:11}.

We then demonstrate an easy to follow method for the design and construction of a  wideband homodyne detector using commercial available components such as low-noise high-speed operational amplifiers and high-bandwidth photodiodes. Aside from high performance benchmarks, a special feature of our detector is its versatility: it is designed and tested to operate in both the continuous-wave or pulsed regimes. Therefore the unit presented here may be useful for a wide range of quantum optics experiments.


\section{Theoretical analysis}
Balanced homodyne detection consists of overlapping the signal mode carrying the quantum state in question and a strong reference field in a matching mode (the local oscillator, or LO) on a symmetric beam splitter. The two output signals of this beam splitter are directed to the two photodiodes of the HD, where these fields are detected and subtracted. Neglecting experimental imperfections, the subtraction photocurrent is then
\begin{equation}\label{idealHD}
\hat i(t)=A\alpha(t)\hat q_\theta(t),
\end{equation}
where $\hat{q}_\theta(t)$ is the instantaneous field quadrature value in the signal mode, $\alpha(t)$ and $\theta$ are the local oscillator amplitude and phase, respectively, and $A$ is a proportionality coefficient related to the HD amplifier gain. It is assumed that the local oscillator phase is constant.

The instantaneous quadrature observable can be written as
\begin{equation}\label{qtheta}
\hat{q}_\theta(t)=\hat a(t)e^{i\theta}+\hat a^\dag(t)e^{-i\theta},
\end{equation}
where $\hat a(t)$ is the time dependent photon annihilation operator\cite{fed:05}.

In a practical HD, the relationship between the quadrature measurement and the output current is more complex. It can be approximated by
\begin{equation}\label{cpracHD}
\hat i(t)=\hat i_e(t)+A\intinf \alpha(t')\hat q(t')r(t-t')\de t',
\end{equation}
where $i_e(t)$ is the detector's electronic noise and $r(\cdot)$ is its response function. An ideal detector would have $i_e(t)=0$ and $r(\tau)=\delta(\tau)$. In practice these conditions are not met.

As evident from \eeqref{cpracHD}, the impact of the electronic noise is minimized by raising the power of the local oscillator and the amplifier gain. However, practical possibilities of increasing the gain without proportionally increasing the electronic noise are limited. The local oscillator power must also be restricted to avoid saturation of the photodiodes and eliminating the classical noise \cite{bac:04}. Therefore in the analysis below we assume $A$ and $\alpha$ to equal their optimal values for the given experimental setting.

\subsection{Continuous regime}
In the continuous regime, the amplitude of the local oscillator is a constant: $\alpha(t)\equiv\alpha$. We are interested in measuring the quadrature of the signal field associated with a particular (normalized) temporal mode function $\phi(t)$, which we assume real:
\begin{equation}\label{QCV}
\hat Q=\intinf \hat q(t)\phi(t)\de t
\end{equation}
To that end, we integrate the homodyne photocurrent with a certain weight function $\psi(t)$, obtaining a \emph{measured} quadrature value,
\begin{eqnarray}\label{QmeasCV}
  \hat Q_{\rm meas}&=&\intinf \hat i(t)\psi(t)\de t \\ \nn
  &=&A\alpha\intinf\intinf \hat q(t')\psi(t)r(t-t')\de t \de t'+\hat Q_e,
\end{eqnarray}
in which the last term,
\begin{equation}\label{}
\hat Q_e=\intinf \hat i_e(t)\psi(t)\de t,
\end{equation}
corresponds to the electronic noise contribution, which we will discuss later. First, we discuss the effect of finite detector response function (bandwidth) on the quadrature measurement.

Equation \eqref{QmeasCV} can be rewritten as
\begin{equation}\label{QmeasCV1}
\hat Q_{\rm meas}=A\alpha\intinf \hat q(t')\psi'(t')\de t'+\hat Q_e,
\end{equation}
where
\begin{equation}\label{}
\psi'(t')=\intinf \psi(t)r(t-t')\de t.
\end{equation}
By comparing Eqs.~\eqref{QCV} and \eqref{QmeasCV1} we find that, by choosing $\psi(t)$ such that $\psi'(t)=\phi(t)$, we have $\hat Q_{\rm meas}=A\alpha\hat Q+\hat Q_e$, i.e., the distortions associated with the detector's finite bandwidth are completely eliminated. This may however be difficult in practice, because the required weight function is a \emph{deconvolution} of the temporal mode of interest and the detector's response. Lack of precise knowledge of either of the above may lead to significant errors in deconvolving.

\begin{figure}[h]
\begin{center}
\includegraphics[width=0.8\columnwidth]{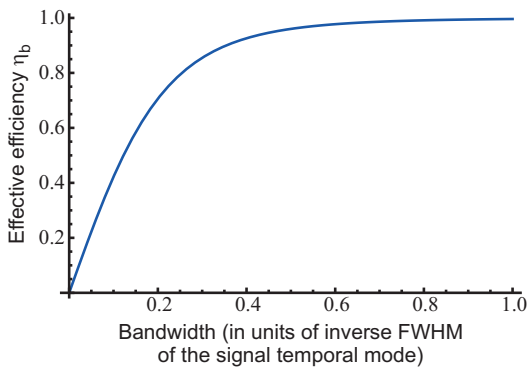}
\caption{Effective efficiency \eqref{MMCV} of the HD associated with its finite bandwidth. The 3-dB bandwidth is plotted along the horizontal axis in units of the inverse full-width-at-half-maximum of the signal temporal mode $\phi(t)$. Both functions are assumed Gaussian. \label{MMCVfig}}
\end{center}
\end{figure}

If $\psi'(t)\ne\phi(t)$, the detection efficiency is degraded by the mode matching factor \cite{aic:02}
\begin{equation}\label{MMCV}
\eta_b=\left.\left|\intinf\psi'(t)\phi(t)\de t\right|^2\middle/\intinf|\psi'(t)|^2\de t\right.,
\end{equation}
where the denominator normalizes $\psi'(t)$. A practically important particular case is when the temporal mode of the signal is known and the finite bandwidth of the detector is neglected, so $\psi(t)$ is set to equal $\phi(t)$. In Fig.~\ref{MMCVfig}, the efficiency \eqref{MMCV} obtained in this setting is plotted for Gaussian $\phi(t)$ and $r(t)$ as a function of the detector bandwidth, which, as we show below, is obtained from the Fourier transform of the response function. 
As we see, the detector bandwidth has no significant degrading effect on the measurement ($\eta_b>0.99$) as long as it is comparable to or larger than the inverse temporal width of the signal temporal mode.

Let us now calculate the contribution of the electronic
noise to the measured quadrature, so the equivalent efficiency \eqref{etae} can be estimated. We start with the electronic spectra of the HD output photocurrent in the absence and in the presence of the local oscillator, with the signal in the vacuum state. According to the Wiener-Khintchine theorem, these spectra are given, respectively, by
\begin{equation}\label{}
S_e(\nu)=\intinf \langle \hat i_e(t)\hat i_e(t+\tau)\rangle e^{2\pi i\nu\tau}\de\tau
\end{equation}
and
\begin{equation}\label{Snu}
S(\nu)=\intinf \langle \hat i(t)\hat i(t+\tau)\rangle e^{2\pi i\nu\tau}\de\tau,
\end{equation}
where $\nu$ is the frequency and the averaging is performed over both time $t$ and the quantum ensemble of the vacuum signal state.
The autocorrelation function in the latter equation can be further simplified as
\begin{eqnarray}
\langle \hat i(t)\hat i(t+\tau)\rangle&& \\\nn &&\hspace{-1.8cm}
= A^2\alpha^2 \left\langle\intinf\intinf \hat q(t')r(t-t')\hat q(t'')r(t+\tau-t'')\de t'\de t''\right\rangle \\\nn
&&\hspace{-1cm}+\langle \hat i_e(t)\hat i_e(t+\tau)\rangle\\\nn
&&\hspace{-1.8cm}=A^2\alpha^2\intinf\langle r(t-t')r(t+\tau-t')\rangle\de t'+\langle \hat i_e(t)\hat i_e(t+\tau)\rangle\\\nn &&\hspace{-1.8cm}=A^2\alpha^2\intinf r(t)r(t+\tau)\de t+\langle \hat i_e(t)\hat i_e(t+\tau)\rangle
\end{eqnarray}
because, in the vacuum state, $\langle \hat q(t')\hat q(t'')\rangle=\delta(t'-t'')$. From the above, we find
\begin{equation}\label{SnuCV}
S(\nu)=A^2\alpha^2|\tilde r(\nu)|^2+S_e(\nu),
\end{equation}
with
\begin{equation}\label{}
\tilde r(\nu)=\intinf r(t) e^{2\pi i \nu\tau}\de t
\end{equation}
being the Fourier image of $r(t)$. In other words, neglecting the electronic noise, the spectrum of the HD output current in the continuous regime is simply the squared amplitude of the Fourier transform of the detector's response function. Note, however, that the response function cannot be obtained from this spectrum because inverse Fourier transformation also requires data on the phase of $\tilde r(\nu)$. The response function can be measured directly in the time domain with a pulsed LO as discussed below.

 Let us now discuss the contribution of electronic noise to quadrature measurements. From \eeqref{QmeasCV1}, and because in the vacuum state $\langle \hat q(t)\hat q(t')\rangle=\delta(t-t')$, we can write
\begin{eqnarray}\label{QmeasCVvar}
  \langle \hat Q_{\rm meas}^2\rangle &=& A^2\alpha^2\intinf |\psi'(t)|^2\de t + \langle \hat Q_e^2\rangle\\ \nn
  &=& A^2\alpha^2\intinf |\tilde\psi'(\nu)|^2\de \nu+ \langle \hat Q_e^2\rangle\\ \nn
  &=& A^2\alpha^2\intinf |\tilde\psi(\nu)|^2|\tilde r(\nu)|^2\de \nu+ \langle \hat Q_e^2\rangle,
\end{eqnarray}
where the variance of the electronic noise component is given by
\begin{eqnarray}
  \langle \hat Q_e^2\rangle
  &=& \intinf\intinf \langle \hat i_e(t)\hat i_e(t+\tau)\rangle\psi(t)\psi(t+\tau)\de t\de\tau\nn \\
  &=& \intinf S_e(\nu)|\tilde\psi(\nu)|^2\de\nu.\label{QeCV}
\end{eqnarray}
Combining the above two results with Eq.~\eqref{SnuCV}, we find
\begin{equation}\label{QmeasCVvar1}
  \langle \hat Q_{\rm meas}^2\rangle = \intinf S(\nu) |\tilde\psi(\nu)|^2 \de\nu.
\end{equation}
Equations \eqref{QeCV} and \eqref{QmeasCVvar1} lead us to an important conclusion: by knowing the homodyne output spectra $S(\nu)$ and $S_e(\nu)$, as well as the weight function $\psi(t)$, one can predict the fraction of the electronic noise in the measured quadrature variance in an arbitrary temporal mode. This, as discussed above, directly translates into an equivalent optical loss.

In the case of a high-bandwidth detector, when $S(\nu)$ and $S_e(\nu)$ can be assumed constant over the support of $\tilde\psi(\nu)$, we have
\begin{equation}\label{}
1-\eta_e\approx \left[\frac {S_e(\nu)}{S(\nu)}\right]_{\textrm{signal bandwidth}}.
\end{equation}
This quantity, which we call \emph{clearance} of the detector's shot noise over the electronic noise, is one of the primary characteristics of any HD circuit.

\subsection{Pulsed regime}
Now let us suppose that the LO is pulsed, with the pulse width much shorter than the time resolution of the electronics. In this case, \eeqref{cpracHD} takes the form
\begin{equation}\label{pracHD}
\hat i(t)= A\alpha_p\hat Q r(t)+\hat i_e(t),
\end{equation}
where the pulse is assumed to occur at $t=0$, $\hat Q=\intinf \alpha(t) \hat q(t)\de t/\alpha_p$ is the normalized  quadrature operator corresponding to the signal mode defined by the shape of the LO pulse, with 
$\alpha_p=\sqrt{\intinf|\alpha(t)|^2\de t}$ being the effective amplitude of the local oscillator pulse. In other words, neglecting the electronic noise, the shape of the HD response to a single short pulse is given by the detector's response function. 

The quadrature measurement is obtained by integrating the homodyne photocurrent over a certain time interval:
\begin{equation}\label{Qmeas}
\hat Q_{\rm meas}=\int\limits_{t_1}^{t_2}\hat i(t)\de t=A\alpha_p \hat Q\int\limits_{t_1}^{t_2}r(t)\de t+\hat Q_e,
\end{equation}
with $$\hat Q_e=\int\limits_{t_1}^{t_2}\hat i_e(t)\de t.$$ The optimal choice of the integration limits is determined by the bandwidths of the detector's electronic noise and the temporal width of its response function. 

\begin{figure}[h]
\begin{center}
\includegraphics[width=0.8\columnwidth]{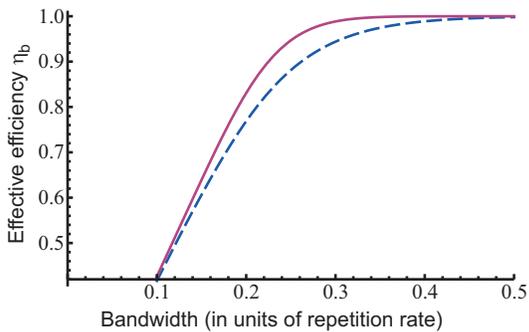}
\caption{Effective efficiency of the HD associated with temporal overlap of responses to different pulsed modes, as a function of the 3-dB bandwidth of the electronics. The response function is assumed Gaussian. The solid line corresponds to a short integration interval $[-\epsilon,+\epsilon]$; the dashed line to the integration interval of length $T$. Both integration intervals  are centered at the peak of the response function. \label{pulsedefftheory}}
\end{center}
\end{figure}

When the local oscillator is a train of pulses with repetition period $T$, the HD output current is given by
\begin{equation}\label{pracHDrep}
\hat i(t)= A\alpha_p\sum\limits_{j=-\infty}^{\infty}\hat Q_j r(t-jT)+\hat i_e(t),
\end{equation}
where $\hat Q_j$ is the quadrature operator of the $j$th pulsed signal mode, with the pulse of interest having index $j=0$. If the response function is nonzero over an interval longer than $T$, the quadrature measurement is contaminated by that of the neighboring pulses:
\begin{equation}\label{QmeasSum}
\hat Q_{{\rm meas},0}=A\alpha_p \sum\limits_{j=-\infty}^{\infty}R_j\hat Q_j +\hat Q_e, \end{equation}
where $R_j=\int\limits_{t_1}^{t_2}r(t-jT)\de t$. The sum in \eeqref{QmeasSum} defines a new measured mode whose state is not necessarily identical to that in the $j=0$th pulsed mode. The corresponding mode matching efficiency (neglecting the electronic noise) is given by
\begin{equation}\label{}
\eta_b=\frac{R_0^2}{\sum\limits_{j=-\infty}^{\infty}R_j^2}.
\end{equation}
This efficiency is plotted in Fig.~\ref{pulsedefftheory} for the response function of Gaussian shape 
as a function of the 3-dB bandwidth of the detector response function spectrum. As we see, a detector bandwidth of at least $~0.4/T$ is required for $\eta_b$ to exceed 99\%.

If $r(t)$ is known, so are all $R_j$ and the effect of finite bandwidth can be reversed by means of discrete deconvolution, akin to the continuous case. However, partial reversal can be implemented even if the response function is not known, provided that $\eta_b$ is sufficiently high, i.e. $|R_j|\ll|R_0|$ for $j\ne 0$, as follows. In a typical detector, $R_j$ are negligibly small for $j>0$: most of the ringings occur \emph{after} the optical pulse that generates the response. This is the situation, for example, with the detector assembled in this work.
Then we have, according to \eeqref{QmeasSum}, and because $\langle\hat Q_j\hat Q_k\rangle=\delta_{jk}$ in the vacuum state,
\begin{equation}\label{}
\langle \hat Q_{\rm meas,0}\hat Q_{{\rm meas},i}\rangle=A^2\alpha^2_p\sum\limits_j R_jR_{j-i}\approx A^2\alpha^2_p R_0R_{-i},
\end{equation}
where $\hat Q_{{\rm meas},i}$ denotes the quadrature measurement for the $i$th pulse. The above correlation can be easily obtained experimentally, from which one can determine
\begin{equation}\label{}
\frac {R_{-i}}{R_0}\approx \frac{\langle \hat Q_{\rm meas,0}\hat Q_{{\rm meas},i}\rangle}{\langle \hat Q_{\rm meas,0}\hat Q_{{\rm meas},0}\rangle}.
\end{equation}
One then calculates
\begin{equation}\label{}
Q'_{{\rm meas},0}=Q_{{\rm meas},0}- \sum\limits_{i=1}^{\infty}\frac{\langle \hat Q_{\rm meas,0}\hat Q_{{\rm meas},i}\rangle}{\langle \hat Q_{\rm meas,0}\hat Q_{{\rm meas},0}\rangle} Q_{{\rm meas},i}
\end{equation}
for each experimentally measured quadrature, thereby eliminating contamination from neighboring pulses. The resulting quadrature values are then renormalized and used in quantum state reconstruction. By means of this technique, the reconstruction efficiency has been improved in Ref.~\cite{hui:09}.

We now use \eeqref{Snu} to determine the spectral power of the HD output in the pulsed regime.
The ensemble average of $i(t)i(t+\tau)$ is a periodic function of time $t$, hence we can write
\begin{eqnarray}\label{}
\langle \hat i(t)\hat i(t+\tau)\rangle&=&\frac 1T (A\alpha_p)^2\\ \nn &&\hspace{-2cm}\times\sum\limits_{j,k=-\infty}^{\infty}\langle\hat Q_j\hat Q_k\rangle\int\limits_{-T/2}^{T/2}r(t-jT)r(t+\tau-kT)\de t\\ \nn&&\hspace{-2cm}+\langle \hat i_e(t) \hat i_e(t+\tau)\rangle_t.
\end{eqnarray}
In the vacuum state,
\begin{equation}\label{}
\langle \hat i(t)\hat i(t+\tau)\rangle=\frac 1T (A\alpha_p)^2\intinf r(t)r(t+\tau)\de t+\langle \hat i_e(t) \hat i_e(t+\tau)\rangle_t.
\end{equation}
Accordingly,
\begin{equation}\label{Sm}
S(\nu)=\frac 1T (A\alpha_p)^2 |\tilde r(\nu)|^2+S_e(\nu).
\end{equation}
We see that in spite of the pulsed character of the local oscillator, the HD spectrum is determined by the Fourier transform of its response function akin to the continuous case. An important difference is the multiplication by the pulse repetition rate: when the separation between the pulses is increased, the spectral power reduces proportionally.

In contrast to the continuous regime, in the pulsed case the evaluation of the equivalent efficiency \eqref{etae} requires knowledge of $r(t)$; information on the spectra $S(\nu)$ and $S_e(\nu)$ is not sufficient. Let us, however, consider a practically important particular case when the bandwidth of both the detector's response and the electronic noise greatly exceed the laser repetition rate. Suppose the integration in \eeqref{Qmeas} is done over the time interval $[-T_0/2,T_0/2]$ with $T_0<T$. Then we have
\begin{equation}\label{Qmeas2}
\langle \hat Q_{\rm meas}^2 \rangle=\left\langle \left(A\alpha_p\hat Q\int\limits_{-T_0/2}^{T_0/2}r(t)\de t\right)^2\right\rangle+\langle \hat Q_e^2\rangle.
\end{equation}
We assume that the temporal width of the function $r(t)$ is much less than $T_0$, so the integration limits can be replaced by $\pm\infty$. We then have
\begin{equation}\label{Qmeas2hb}
\langle \hat Q_{\rm meas}^2 \rangle=(A\alpha_p)^2|\tilde r(0)|^2+\langle \hat Q_e^2\rangle,
\end{equation}
where
\begin{eqnarray}\label{Qe2hb}
  Q_e &=&  \left\langle \left(\int\limits_{-T_0/2}^{T_0/2}\hat i_e(t)\de t\right)^2\right\rangle \\ \nn
  &=& \left\langle \int\limits_{-T_0/2}^{T_0/2}\int\limits_{-T_0/2}^{T_0/2}\hat i_e(t_1)\hat i_e(t_2)\de t_1\de t_2\right\rangle\\ \nn
  &=&  \int\limits_{-T_0/2}^{T_0/2}\int\limits_{-T_0/2-t_1}^{T_0/2-t_1}\langle \hat i_e(t_1)\hat i_e(t_1+\tau)\rangle\de \tau\de t_1\\ \nn
  &\approx&  \int\limits_{-T_0/2}^{T_0/2}\int\limits_{-\infty}^{+\infty}\langle \hat i_e(t_1)\hat i_e(t_1+\tau)\rangle\de \tau\de t_1\\ \nn
  &=&  T_0 S_e(0).
\end{eqnarray}
Here we again took advantage of the high bandwidth of the electronic noise to modify the integration limits. We also used the fact that the electronic noise $i_e(t)$ is a stationary stochastic process to eliminate the dependence on $t_1$. Substituting Eqs.~\eqref{Sm}--\eqref{Qe2hb} into \eeqref{etae}, we find
\begin{equation}\label{etaeS}
 \eta_e=1-\frac {T_0} T \frac{\hat S_e(0)}{ S(0)}.
\end{equation}
In other words, the shot-to-electronic-noise clearance measured in the pulsed regime yields a too conservative estimate for the equivalent loss associated with the electronic noise in the high-bandwidth limit. In order to minimize this loss, one needs to choose the integration interval that is as short as possible, but still accommodates the entire detector response function.
However, if the detector bandwidth does not greatly exceed the laser repetition rate, we have $T_0\simeq T$ and the factor of $T_0/T$ in the above equation can be neglected. The exact value for the equivalent loss in this case cannot be determined from the spectra because it depends on the shape of the detector's response function $r(t)$.

\section{Design and challenges}\label{designSec}
Now we present an efficient wideband HD constructed with easily available components, as well as tips to solve the most common challenges found when building such a device. We show the results given by our HD in experiments with a pulsed laser source (76 MHz repetition rate) as well as with a continuous wave laser. These features give our HD a great versatility for applications in different kinds of experiments.

\begin{figure*}[t]
\begin{center}
\includegraphics[scale=0.65]{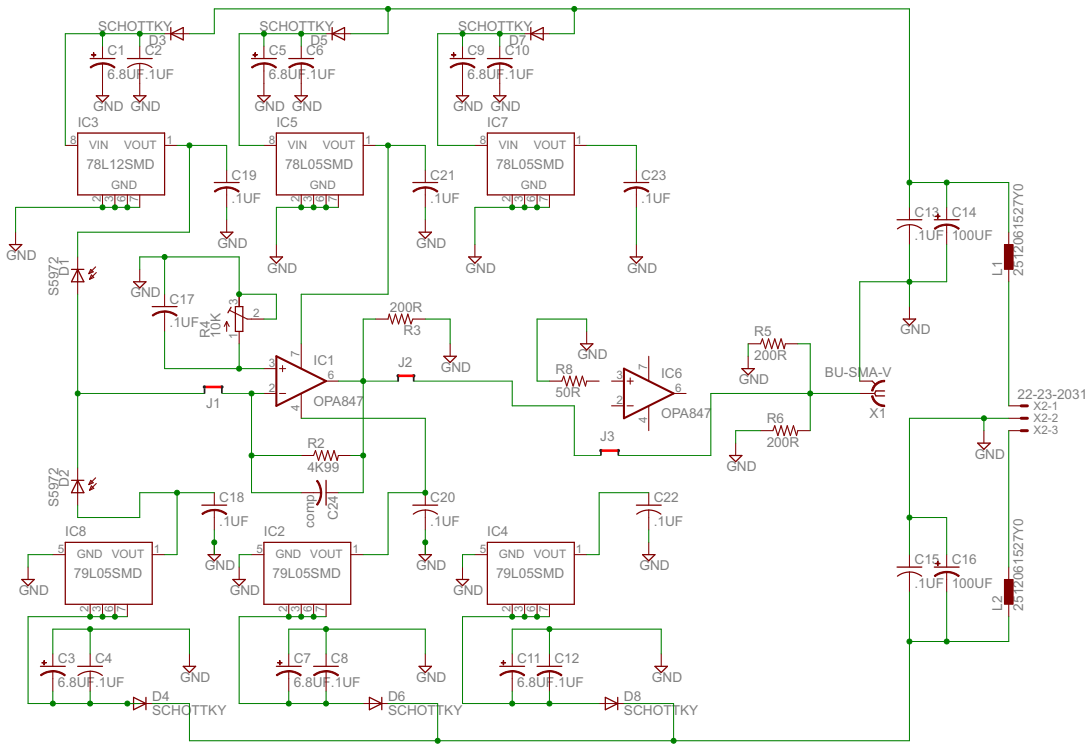}
\caption{HD Circuit diagram\label{fig:schm}}
\end{center}
\end{figure*}

Our electronic cicruit is shown in Fig.~\ref{fig:schm}. The values and identification names are as noted on the schematic. Provision has been made for a second stage of voltage gain which, however, proved to be unnecessary. 

We use a homemade $\pm 15$ V DC power supply. To avoid possible leakage of any ambient noise to the HD through the voltage supply lines, each of them is filtered for both high and low frequencies (components L1, L2, C13, C14, C15, C16 that include a ferrite choke, a 100 $\mu$f electrolytic and a 0.1 $\mu$f ceramic capacitors). The use of long power supply lines in the circuit can create parasitic inductances making the circuit susceptible to instabilities. Because it is not possible to make all the tracks short, we perform the filtering in two stages, at the beginning and the end of each power supply track.

The filtered $\pm 15$ Volt supply rails are each further separated into 3 isolated lines by incorporating 3 networks consisting of a fast Schottky diode (for improving the stability of the power supplies), a tantalum 6.8 $\mu$f capacitor and a 0.1 $\mu$f ceramic capacitor (components C1--C12 and D3--D8). These isolated supplies in turn feed $\pm 12$ V regulators for the photodiodes (IC3 and IC8) and two pairs of $\pm 5$ V regulators for the first and second stage operational amplifiers (IC2, IC5, IC4 and IC7). Care has been taken to isolate and regulate the supply voltage for the detectors and amplifiers in order to provide as stable and oscillation free platform as possible. Bypass 0.1 $\mu$f capacitors are located at the regulator outputs very close to the amplifiers (C18--C23).

One of the primary challenges in achieving a high-bandwidth HD is associated with the photodiodes. Their P-N junction has a terminal capacitance which has to be significant for high quantum efficiency of detection. The feedback resistance of the amplifier will form a low pass filter with the terminal capacitance, thus limiting the high-frequency response of the circuit. Furthermore, this capacitance combined with other capacitances and inductances in the circuit board can  give rise to instabilities and oscillations associated with a low pass filter configuration.

The terminal capacitance of the photodiodes can be reduced by increasing the reverse bias voltage across them. This, however, increases the dark current that contributes to the electronic noise \cite{zav:02}. Consequently, special care must be taken for choosing photodiodes with very low capacitance and dark current, as well as for a printed circuit board (PCB) design that avoids or compensates any capacitance that could produce oscillations (instabilities) in the frequency response of the HD.

We have chosen to use Hamamatsu S5972 photodiodes. This photodiode has a 91\% quantum efficiency at a wavelength of 780 nm and a high cut-off frequency (500 MHz) when supplied in a 12 V reverse voltage configuration. Other reasons for this choice include a very low dark current (11 pA at 12V reverse voltage) and a low terminal capacitance (2.8 pf at 12V reverse voltage).

The amplification circuit starts at the junction of the two S5972 photodiodes, which is the differential sum point for the input to the first stage amplifier. The amplification of the difference signal is carried out by a single Texas Instruments OPA847 operational amplifier (op-amp) in trans-impedance configuration with a trans-impedance gain of 4 k$\Omega$ (IC1). For testing purposes we place a jumper (J1) in the inverting input of the op-amp. In order to have control of the DC offset of output signal produced by the the op-amp we place in its non-inverting input a capacitor of 0.1 $\mu$f in parallel with a variable resistor of 10 k$\Omega$ both connected to ground (C17 and R4). With the help of the variable resistor we can maintain the DC offset at the zero reference point to avoid saturation in the op-amp. The optional second stage is designed to use an OPA847 in inverting amplifier configuration. J2 and J3 are the positions of its input and feedback resistances.

An important component of the circuit is a custom made variable capacitor (C24) placed in parallel to the feedback resistance R2. This capacitor is constructed by twisting two wires with a variable number of twists. By twisting (or untwisting) these wires we can vary their capacitance; although it is very small,  it is enough to change the response of the circuit. In our case it is used to flatten the spectral response of the HD.

Finally, resistors R3, R5 and R6 are used for impedance matching of the output signal of the amplifier with a 50$\Omega$ coaxial connector.

\begin{figure*}[t]
\begin{center}
\includegraphics[width=6in]{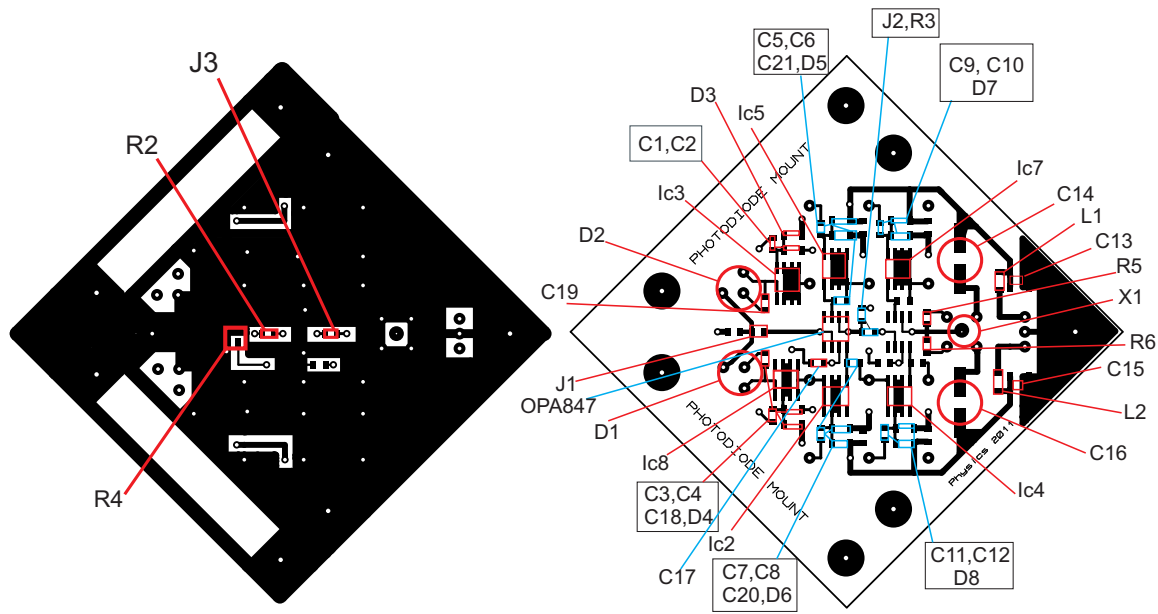}
\caption{HD Layout: (Left) Bottom layer where the ground plane is; (Right) Upper layer, connection tracks and OPAmps\label{fig:pcb}}
\end{center}
\end{figure*}

All components are soldered on a specially designed printed circuit board (Fig.~\ref{fig:pcb}). When designing the layout for the circuit we take the following factors into consideration.
\begin{itemize}
\item The tracks on the PCB are kept as short as possible to avoid parasitic inductances.
\item A ground (GND) plane, essential for high frequency circuits, and a double-sided board design are used. The back side of the PCB is mainly reserved for the GND plane, with the fewest possible discontinuities. This separation between tracks and components in one side and GND on the other minimize possible capacitances between the tracks and the ground plane of the circuit.
\item The use of surface mount components is preferred for the high frequency regime of electronics.
\item In order to provide a mechanically stable platform for the HD, the circuit module is mounted on a 0.25-inch aluminum plate.
\item To ensure a maximally balanced response, we designed our circuit to be as symmetrical as possible.
\item The HD is shielded by a custom-made metal box to avoid any environmental noise impact on the circuit.
\item To minimize the parasitic capacitance the two photodiodes were placed closely together in the upper left corner of the board.
\item Due to its relatively large size, the offset adjustment potentiometer (R4) for the first stage is placed at the bottom of the printed circuit board to avoid the issue of having long tracks on the upper side.
\end{itemize}

\begin{figure}[h]
\begin{center}
\includegraphics[width=2.5in]{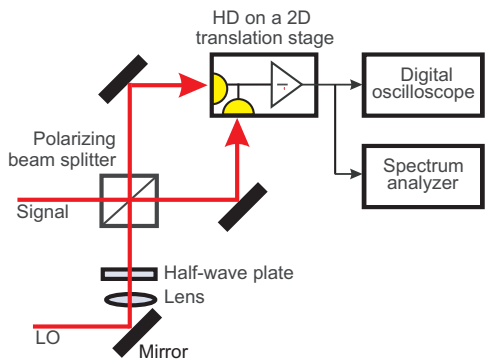}
\caption{Balanced HD experimental set-up\label{fig:setup}}
\end{center}
\end{figure}

\section{Performance and characterization}
\subsection{Pulsed regime}\label{pusedSec}

After having built the HD we proceed to test its  performance. To that end, we place it in the experimental set-up shown in Fig.~\ref{fig:setup}. Our LO is obtained from a mode-locked Coherent MIRA 900 Ti:Sapphire laser producing 1.8 ps pulses with a repetition rate of 76  MHz, central wavelength of 791 nm. The HD beam splitter configuration is implemented by using a half wave plate and a polarizing beam splitter. The two beams coming out of the beam splitter are focused into the photodiodes.

The HD is balanced by first adjusting the waveplate in order to equalize the responses of the photodiodes, compensating for a possible difference in their quantum efficiencies, thereby minimizing the spurious signal at the local oscillator repetition rate. A further crucial step in balancing the HD is to equalize the path lengths of the two beams entering the photodiodes by using a XY translation stage on which the HD is mounted. This is necessary because the pulses, if arriving at the photodiodes at different times, can lead to subtraction pulse having a bipolar shape. Even after these alignment steps, the subtraction may not be perfect due to different capacitances of the two photodiodes.

Subsequently, the HD is tested for electronic instabilities and oscillations. This is done by observing the output signal of the HD (with the LO power of 6 mW) with a spectrum analyzer in a range from 100 kHz up to 3 GHz. If present, instabilities produce peaks in the spectral response at frequencies different from the repetition rate of the LO and its harmonics. They can be removed by adjusting the values of resistances R2, R3, R5 and R6 as well as the custom made variable capacitor C24. Additional instability sources can be associated with the ambient noise or leakage through the power supply lines; in this case, the elements of the power supply filtering sections must be changed.

\begin{figure}[t]
\begin{center}
{
\includegraphics[width=0.9\columnwidth]{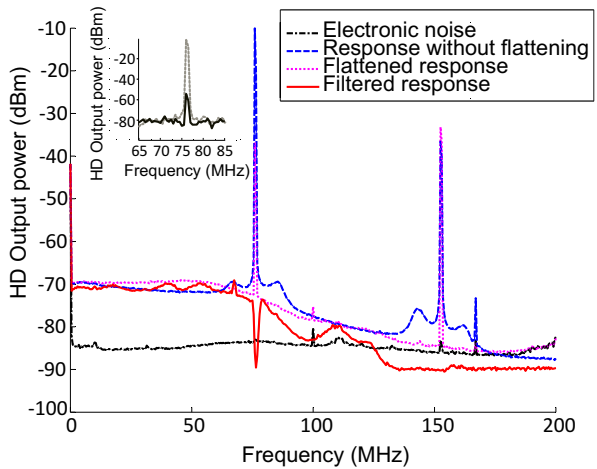}
}
\caption{Spectral response of the HD at 12 mW of LO power. The large peaks are the laser repetition rate and its second harmonic. \emph{Inset}: The response at 100 $\mu$W of LO power, necessary to calculate the CMRR. Grey (top) trace: one photodiode blocked; Black (bottom) trace: both photodiodes illuminated\label{fig:spectralresponsepulse}}
\end{center}
\end{figure}

In the final step of adjustments we flatten the spectral response of the HD. This is done by changing the value of the capacitance C24 while observing the spectrum of the HD output. Fig.~\ref{fig:spectralresponsepulse} (dashed and dotted lines) shows the effect of this adjustment in the spectral profile of the HD.

In order to characterize the detector, we first measure the shot noise clearance. To this end, we increase the LO power to 12 mW. The choice of this power level is determined by the need to achieve the highest shot-to-electronic noise clearance, while at the same time ensuring stable and saturation-free operation of the HD. With the given wavelength and repetition rate, this LO power corresponds to $6.2 \times 10^{8}$ photons per pulse, and will produce $5.6 \times 10^{8}$ photoelectrons per pulse, which corresponds to a shot noise of $24,000$ electrons per pulse. In Fig.~\ref{fig:spectralresponsepulse} (dotted vs. dash-dotted lines) we show the clearance between electronic noise and shot noise, which has a value of 13dB, correponding to $\eta_e=0.95$. The 3-dB bandwidth of the HD is 80 MHz, which implies that it can be applied, without significant efficiency loss, to pulsed lasers with repetition rates up to 250 MHz (Fig.~\ref{pulsedefftheory}).

\begin{figure}[h]
\begin{center}
\includegraphics[width=0.8\columnwidth]{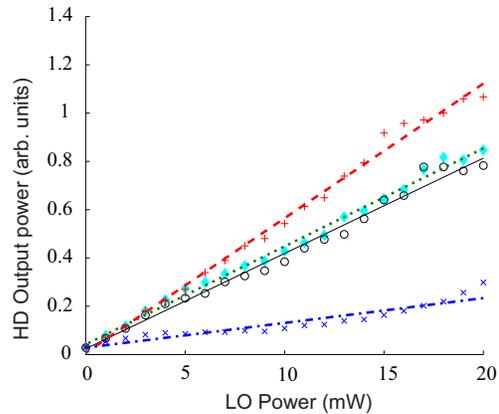}
\caption{Response of the HD respect to the LO power for different fixed frequencies. The linear fits are also shown. Red dash line (plus): 80MHz; Green dot line (diamond): 60MHz; Black continous line (circle): 40MHz; Blue dot dash line (cross): 100MHz.\label{fig:linear}}
\end{center}
\end{figure}

We verify that the power of the observed HD output signal grows linearly with the LO power, which is the signature of the shot noise \cite{bac:04,lvov:09}. In Fig.~\ref{fig:linear} we show measurements of the shot noise level for five different frequencies depending on the LO power. For frequencies up to 100 MHz, the HD behaves linearly, although the slope is reduced for higher frequencies.

In order to determine the CMRR value, we measure the spectral response of the HD at the repetition frequency of the LO at a power of 100 $\mu$W in two cases: when both photodiodes are illuminated, and when only one photodiode is illuminated (inset of Fig.~\ref{fig:spectralresponsepulse}). The measurement is performed with a low LO power to avoid saturation of the HD in the absence of subtraction. We find that our HD has a CMRR of 52.4 dB, which corresponds to a reduction by a factor of $2.4\times10^{-3}$ in the photocurrent corresponding to the common mode, yielding $6.7\times 10^5$ residual photoelectrons per pulse due to imperfect subtraction \cite{chi:11}, which is much greater than the average number of photoelectrons per pulse due to the shot noise.  This is the reason why, even with the optimized alignment, the pulsed HD spectrum exhibits a peak at the  pulse repetition rate.

Theoretically, this residual periodic signal leads to a constant displacement in the quadrature measurement, which is easy to average out. In practice, however, the signal tends to spontaneously change its magnitude every few seconds. Furthermore, the presence of that signal complicates computer acquisition of the HD output because of the limited digitizer resolution. Therefore we use two notch filters (MFC 6367) to eliminate the output components at 76 and 152 MHz.

The HD output signal is further processed by  a 100 kHz high pass filter to remove the DC offset. To clean the signal and to avoid any noise at higher frequencies, two low pass filters with a 100 MHz cut-off are used. The resulting HD spectrum is shown in Fig.~\ref{fig:spectralresponsepulse} (solid line).


\subsection{Continuous regime}
\begin{figure}[h]
\begin{center}
{
\includegraphics[width=0.9\columnwidth]{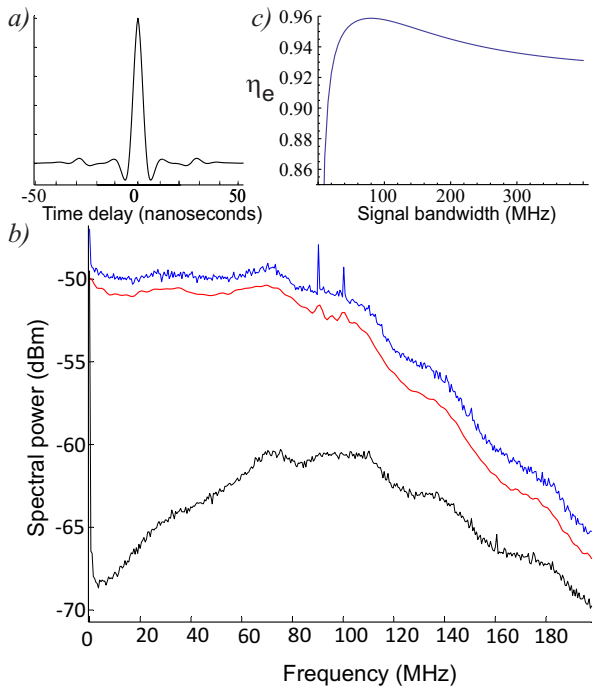}
}
\caption{characterization of the HD in the continuous regime. a) Autocorrelation function. b) Electronic spectrum of the shot noise obtained with a spectrum analyzer (blue trace --- top); the black (bottom) trace is the electronic noise, and by Fourier transform of the autocorrelation function (red trace --- middle). The vertical position on the logarithmic scale of the red trace with respect to the spectrum analyzer traces is chosen arbitrarily. c) Effective efficiency \eqref{etae} associated with the electronic noise as a function of the signal wavefunction bandwidth (in units of the inverse FWHM of its temporal mode). \label{fig:spectralresponsecont}}
\end{center}
\end{figure}

In order to check the versatility of our scheme, we analyze the performance of another HD, constructed similarly to the first one, in the continuous wave regime by using a TekhnoScan Ti:Sapphire laser with a center wavelength of 795 nm and a bandwidth of 5 kHz. We use the same experimental set-up as in Fig.~\ref{fig:setup}. We analyze the  response of the HD in the time domain  by acquiring 30 traces of 100-$\mu$s duration, each containing $2 \times 10^{5}$ points (sampling frequency of 2GHz). Thereafter, we compute the autocorrelation function for each trace, and then average it over the 30 traces [Fig.~\ref{fig:spectralresponsecont}(a)].

We then acquire the HD output spectrum, both in the presence and absence of LO, using a spectrum analyzer. As evidenced by Fig.~\ref{fig:spectralresponsecont}(b), the shot noise spectrum is similar, up to a constant factor, to that obtained by Fourier transform to the autocorrelation function in agreement with the  Wiener-Khintchine theorem. We find this detector to exhibit a 3-dB bandwidth of 100 MHz and the shot-to-electronic noise clearance ranging between 10 and 18 dB.

In Fig.~\ref{fig:spectralresponsecont}(c) we plot the effective loss \eqref{etae} associated with the electronic noise under the assumption that the detector is used to measure field quadratures in a Gaussian temporal mode. The loss is calculated according to Eqs.~\eqref{QeCV}, and \eqref{QmeasCVvar1} given the measured spectra [Fig.~\ref{fig:spectralresponsecont}(b)]. We find that the effective loss strongly varies dependent on the shape and width of the temporal mode.

Table \ref{tab:table1} shows the features of our device in comparison with other HDs reported in the literature. As we see, our detector compares favorably with its counterparts: it shows a unique combination of the bandwidth, CMRR and the electronic noise clearance. 

\begin{table}[h]
\caption{\label{tab:table1}Comparison between various HD's.}
\begin{ruledtabular}
\begin{tabular}{l|r|r|r|r|r|r|r}
Characteristics&Ours&\cite{han:01}&\cite{chi:11}&\cite{zav:02}&\cite{nie:09}&\cite{oku:08}&\cite{had:09}\\
\hline
Wavelength  (nm) &791 &790 &1550 &786 &860 &1064 &--\\
3 dB bandwidth (MHz)&100 &1 &100 &$\sim$50 &100 &250 &54\\
CMRR (dB) &52.4 &85 &46 &42 &-- &45 &61.8\\
Clearance (dB) &13 &14 &13 &$\sim$5 &10 &$\sim$7.5 &12\\
\end{tabular}
\end{ruledtabular}
\end{table}

\section{Summary}

We presented a comprehensive theory and a detailed recipe for designing, building and troubleshooting a wideband balanced detector for highly accurate time-domain quantum measurements. We showed that by following these recommendations a homodyne detector with a shot-noise clearance of 13 dB at 12 mW of local oscillator power, a CMRR value of 52.4dB, a flat response up to 100 MHz can be constructed in an easy way and by only using a single trans-impedance amplification stage. We also have shown that this HD can be applied to both pulsed and CW configurations.

We thank Frank Vewinger, Nitin Jain and Ryan Thomas for helpful discussions. This work has been supported by NSERC, CIFAR, Quantum\emph{Works} and ARC (Grant DP1094650).

\bibliographystyle{aip} 

\end{document}